\documentclass[a4paper]{article}
\usepackage{amsmath,amssymb}
\usepackage[margin=2.5cm]{geometry}

\usepackage{subfigure,color,graphicx}
\begin{document}

\title{Memristive Fingerprints of Electric Arcs}
\author{Wieslaw Marszalek\\
DeVry University, 630 US Highway 1\\
 North Brunswick, NJ 08902, USA\\
email: wmarszalek@devry.edu} 



\maketitle

\begin{abstract}
We discuss the memristive fingerprints of the hybrid Cassie-Mayr model of electric arcs. In particular, it is shown that (i) the voltage-current characteristic of the model has the pinched hysteresis nature, (ii) the voltage and current zero crossings occur at the same  instants, and (iii) when the frequency $f$ of the power supply increases, the voltage-current pinched hysteresis characteristic tends closer to a single-valued one, meaning that the voltage-current graph becomes that of a resistor (with an increased linearity for $f\rightarrow \infty$). The conductance $g$ of the Cassie-Mayr model decreases when the frequency increases. The hybrid Cassie-Mayr model describes therefore an interesting case of a memristive phenomenon.
\end{abstract}


\section{Introduction}

Consider the Cassie-Mayr hybrid model of electric arcs \cite{cass}-\cite{add4}
\begin{equation}\label{1}
g=G_{min}+\left [1-e^{-\frac{i^2}{I_0^2}}\right ]\frac{ui-Ki^2}{U_C^2}+e^{-\frac{i^2}{I_0^2}}\frac{i^2}{P_M}-\theta\frac{dg}{dt}
\end{equation}
driven by a power circuit with the voltage source $E(t)=E_m sin(2\pi ft)$, resistor $R$ and inductor $L$, described by  
\begin{equation}\label{2}
L\frac{di}{dt}+Ri+u=E
\end{equation}
where $u$ and $i$ are the arc voltage and current, respectively, $g$ is the conductance of the arc with $u=i/g$ and $G_{min}$, $I_0$, $K$, $U_C$, $P_M$, $L$, $R$, $E_m$, $f$ are real positive constants, $\theta=\theta_0+\theta_1e^{-\alpha |i|}$, with $\theta_0$, $\theta_1$ and $\alpha$ being constants such that $\theta_0 \ll \theta_1$. When the current $i$ is small, one can consider $\theta \approx\theta _1$, while for large current  $\theta \approx\theta _0$  \cite{01}. Another frequent simplification (however not assumed in this paper) is to have  $K=0$, which  means that no energy dissipation occurs due to plasma radiation.

Also, the positive constant $G_{min}$ plays the role of a minimum value of $g(t)$, as many authors assume that $g(t)> G_{min}$ when the current $i(t)$ is small. The $G_{min}$ is a very small 
conductance between two electrodes when the arc is absent. In general, the value of $G_{min}$ depends on the distance
between the electrodes, their geometry, type of
gas used and temperature. Detailed physical assumptions about the above model can be found, for example, in \cite{01}-\cite{pap7}.

The literature on electric arcs in welding, foundry, gas discharge lamps, lighting as well as voltaic, iron, cobalt, nickel, titunium and mercury arcs is particularly immense over the last 150 years. For example, many papers on electric arcs  were published in the Journal of the Franklin Institute  over a period of more than hundred years - since 1850s to 1950s - see \cite{add0}-\cite{add4} for a few examples of such papers. A list of papers on the topic of electric arcs available  in the literature can really be made impressive and long.

Impressive are also the very recent discoveries in the area of nanotechnology related to memristors and memristive circuits and their properties. The announcement by a group of Hewlett Packard researchers \cite{HP} about a succesful construction of 'the missing memristor' renewed interest in the earlier theoretical work of L. O. Chua and others on memristors \cite{chua1},\cite{chua2}. That research has been significantly expanded in the last few years, see \cite{finger}-\cite{devry} and references therein.

The two seemingly distant areas of electric arcs and memristors are, in fact, close to each other and this paper addresses that  issue through the analysis of the properties of the models of electric arcs and the models of memristors.

In particular, it is shown in this paper that the model (\ref{1}),(\ref{2}) has the three fingerprints of memristors (see \cite{finger},\cite{brain},\cite{lamps}), as follows:
\begin{itemize}
\item The $u$ and $i$ characteristic is of the pinched hysteresis type.
\item The $u$ and $i$ zero crossings occur at the same instants.
\item As $f\rightarrow \infty$, then the $u$-$i$ pinched hysteresis characteristic becomes that of a resistor, meaning that the  $u$-$i$ graph  is a single-valued one with no memory effect.
\end{itemize}

\begin{figure}[t!]
\begin{center}
\subfigure[Pinched hysteresis for $f=50$ Hz.]
{\label{Rys1a}\includegraphics*[height=1.85in,width=3.4in]{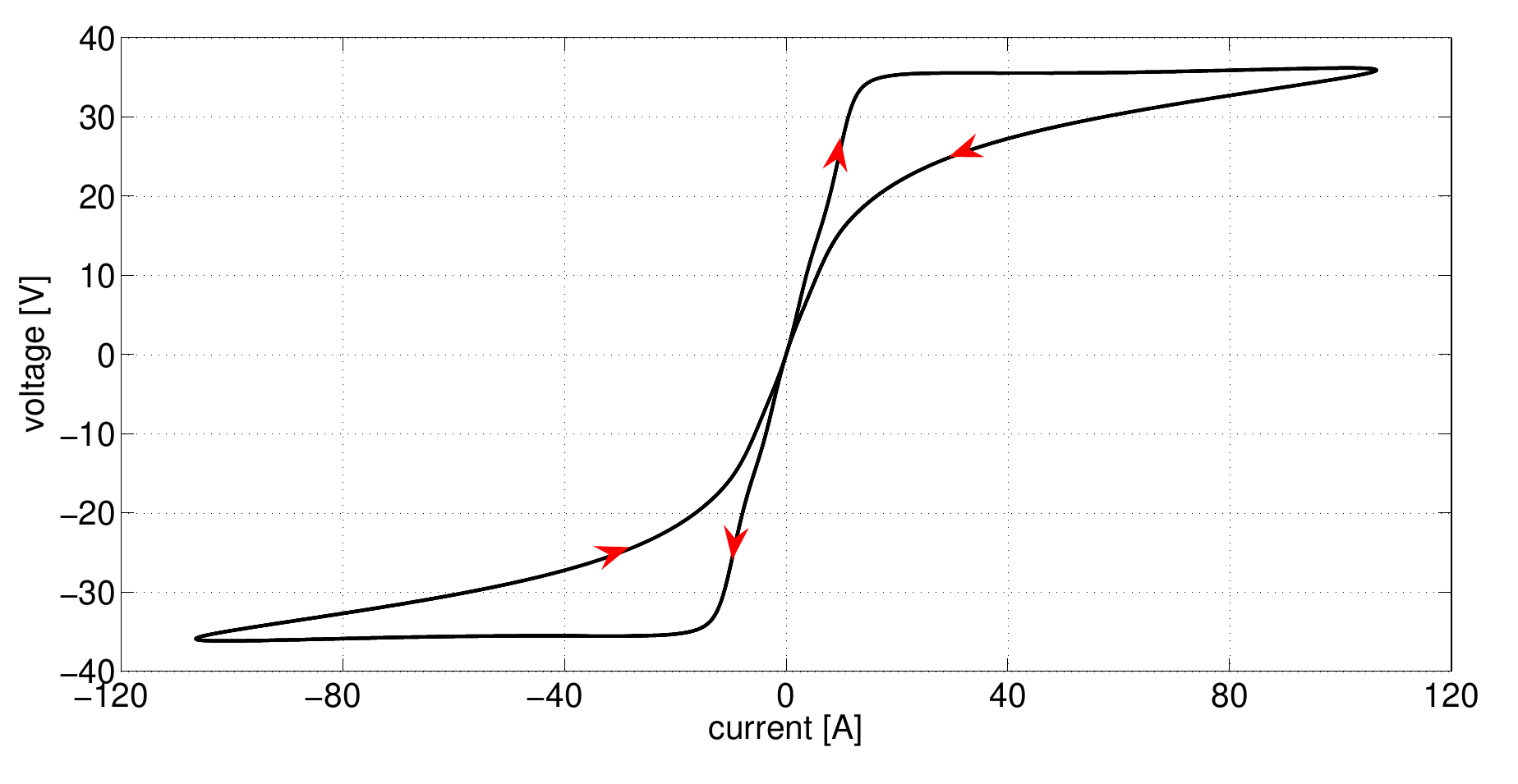}}
\subfigure[Normalized $u(t)$ {\color{red}$\relbar$} and $i(t)$  {\color{blue}$\relbar$} for $f=50$ Hz.]
{\label{Rys1b}\includegraphics*[height=1.85in,width=3.4in]{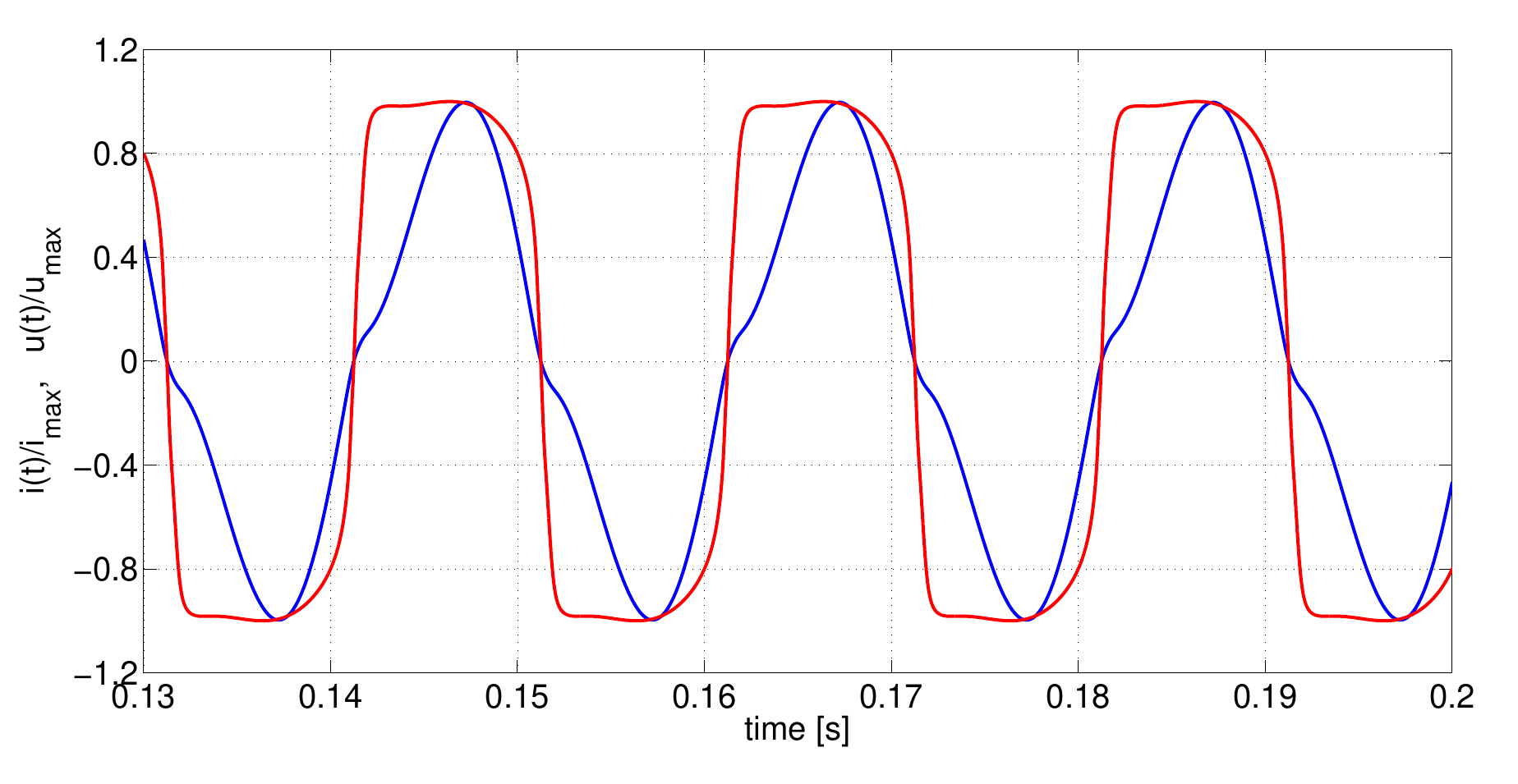}}
\subfigure[Pinched hystereses for $f=\{0.4, 3, 5, 7, 9\}$ kHz.]
{\label{Rys1c}\includegraphics*[height=1.85in,width=3.4in]{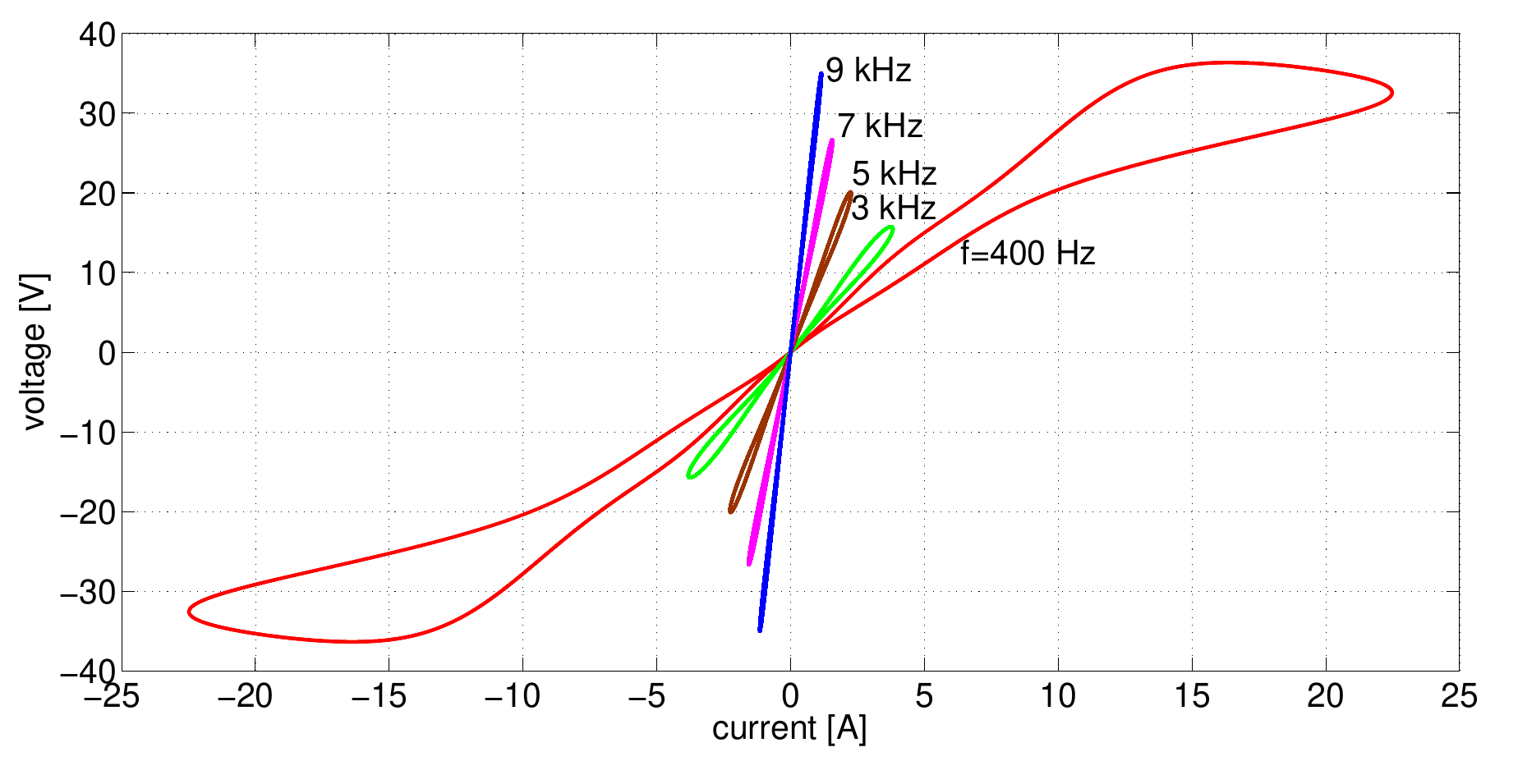}} 
\subfigure[Conductance $g$ versus voltage $u$ for $f=50$ Hz.]
{\label{Rys1d}\includegraphics*[height=1.85in,width=3.4in]{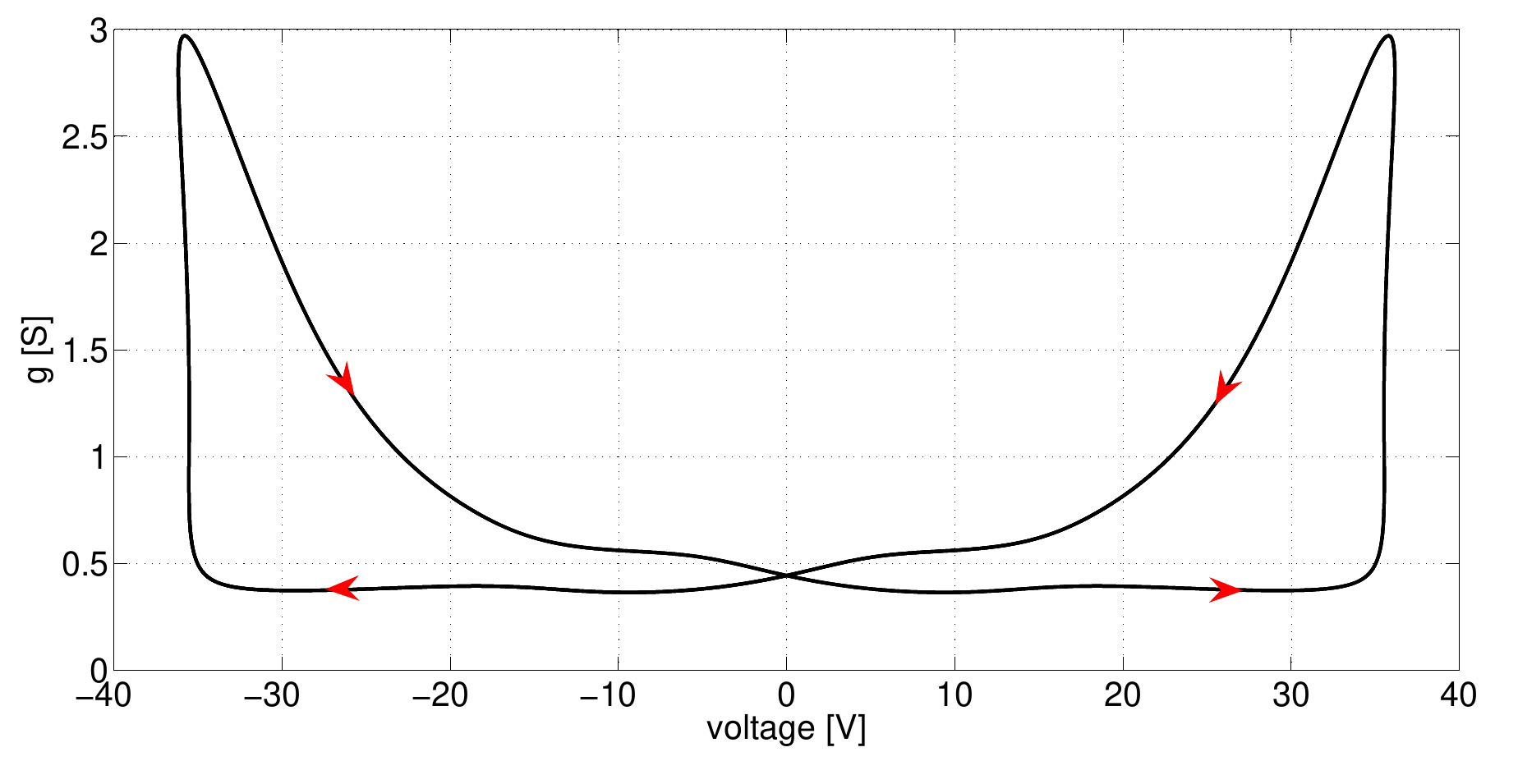}} 
\caption{Solution of (\ref{1}),(\ref{2}) with $u=i/g$. (a)-(c): the three fingerprints of the Cassie-Mayr model; (d) conductance $g$ versus voltage $u$. Parameters $\theta=4~\times~10^{-4}$, $G_{min}=10^{-8}$, $I_0=4.8$, $P_M=20$, $U_C=30$, $R=0.2$, $L=10^{-3}$, $K=10^{-1}$, $E_m=75$. The \emph{ode45} solver from Matlab with \emph{abserr}=\emph{relerr}=$10^{-10}$ was used.}
\end{center}
\label{Fig1} 
\end{figure}

The above fingerprints are illustrated in Fig.\ref{Rys1a}-\ref{Rys1c} for a selected set of constant parameters in (\ref{1}) and (\ref{2}).


\section{Memristors}
 Memristors, as passive elements, complement the widely used other passive elements: resistors, capacitors and inductors. Each of the four passive two-port elements uses a pair of the current, voltage, charge or flux variables as their inputs and outputs. Memristors are nonlinear elements whose present state at any instant depends on the past (i.e. memory). For example, the current-controlled voltage memristor is described by a relationship between the flux $\phi(t)$ and charge $q(t)$, as follows: $\phi(t)=F(q(t))$ with some function $F\in C^1$. This gives the Ohm's law for such a memristor in the form:  $u(t)=r(\int i(t)dt)i(t)$, with $r(q)=dF(q)/dq$, the memristance, while $u(t)$ and $i(t)$ denote  the voltage and current, respectively. The memory effect is due to the dependence of memristance $r$ on $\int i(t)dt$. Other types of mem-elements are also possible (see \cite{Paper10} and  references therein). The recent papers \cite{finger},\cite{brain}-\cite{mech1} show interesting electrical, mechanical and biological devices and phenomena, all having the features (the so-called fingerprints) of memristors. This paper goes in the same direction and proves mathematically that the hybrid Cassie-Mayr model of an electric arc \cite{01} has the three fingerprints of memristors (in the time- and frequency domains).

\vspace{.15cm}
First, it is worth pointing out that the fundamental idea behind the model (\ref{1}),(\ref{2}) is to have the conductance $g(t)$ as a combination of the conductances  $g_C(t)$ and $g_M(t)$, as follows \cite{pap4},\cite{pap8}
\begin{equation}\label{3}
g(t)=\left [1-\sigma(i(t))\right ]g_C(t)+\sigma(i(t))g_M(t)
\end{equation}
where $g_C$ and $g_M$ are the conductances obtained from the Cassie and Mayr models, respectively, and the weighting function $0\le \sigma(i)\le 1$ is monotonically increasing when $i(t)$ increases. Also, typically $\sigma(0)=1$. The most common choice is to have $\sigma(i)=e^{-i^2/I_0^2}$ in (\ref{1}), with $I_0=const$ being the \emph{transition} current. When $i(t)$ is much smaller than $I_0$, the Mayr model is dominant in (\ref{1}), while for $i(t)$ large, the Cassie model dominates in (\ref{1}). This feature of the hybrid Cassie-Mayr model is similar to that of the Hewlett Packard (HP) memristor's model  in which the total memristance is obtained as a weighted sum of $R_{OFF}$ and $R_{ON}$ resistances \cite{HP}

\begin{equation}\label{4}
R(t)=\left [1-\frac{w(t)}{D}\right ]R_{OFF}(t)+\frac{w(t)}{D}R_{ON}(t)
\end{equation}
where $R_{ON}$ and $R_{OFF}$ denote the resistances of the region with a
high concentration of dopants (having low resistance $R_{ON}$), and the region with a low
 dopant concentration (having much higher resistance
$R_{OFF}$), respectively. Thus, $R_{ON}\ll R_{OFF}$ and $0\le w(t)\le D$. As a consequence, we have a one-to-one correspondence between (\ref{3}) and (\ref{4}). Namely, $w(t)\ll D$ gives $R(t)\approx R_{OFF}$ resulting in  a \emph{small} memristor's current.  A \emph{small} $i(t)$, that is $i(t)\ll I_0$  in an electric arc, gives $g(t)\approx g_M$. On the other hand, $w(t)$ close to $D$ yields $R(t)\approx R_{ON}$ and a \emph{large} memristor's current. A \emph{large} $i(t)$, that is $i(t)\gg I_0$ in an arc, results in $g(t)\approx g_C$.

The  $\sigma(i)=e^{-i^2/I_0^2}$ used in (\ref{3}) is not the only possible function used in hybrid arc models. Other monotonical functions $\sigma(i(t))$ used in modeling of electric arcs are $e^{-\left (\frac{|i|}{I_0}\right )^a}$, $e^{-\left (\frac{|i|}{I_0}\right )^{a/(\delta +|i|)}}$ or $1/\left [1+e^{\beta(|i|-I_0)}\right ]$ for constants $a$, $\delta$ and $\beta$ \cite{pap8}.

The above one-to-one correspondence between an electric arc and a memristor is further obvious by analyzing the three fingerprints of memristive phenomena, mentioned above and analyzed in detail in the next section.

\section{The three memristive fingerprints of electric arcs}

\vspace{.1cm}
\noindent {\bf Fingerprint 1:}  The $u$-$i$ characteristic of (\ref{1}),(\ref{2}) is of the pinched hysteresis type.

\begin{figure}[h!]
\begin{center}
\subfigure[Pinched hystereses for different $I_0$ values.]
{\label{Rys2a}\includegraphics*[height=1.85in,width=3.4in]{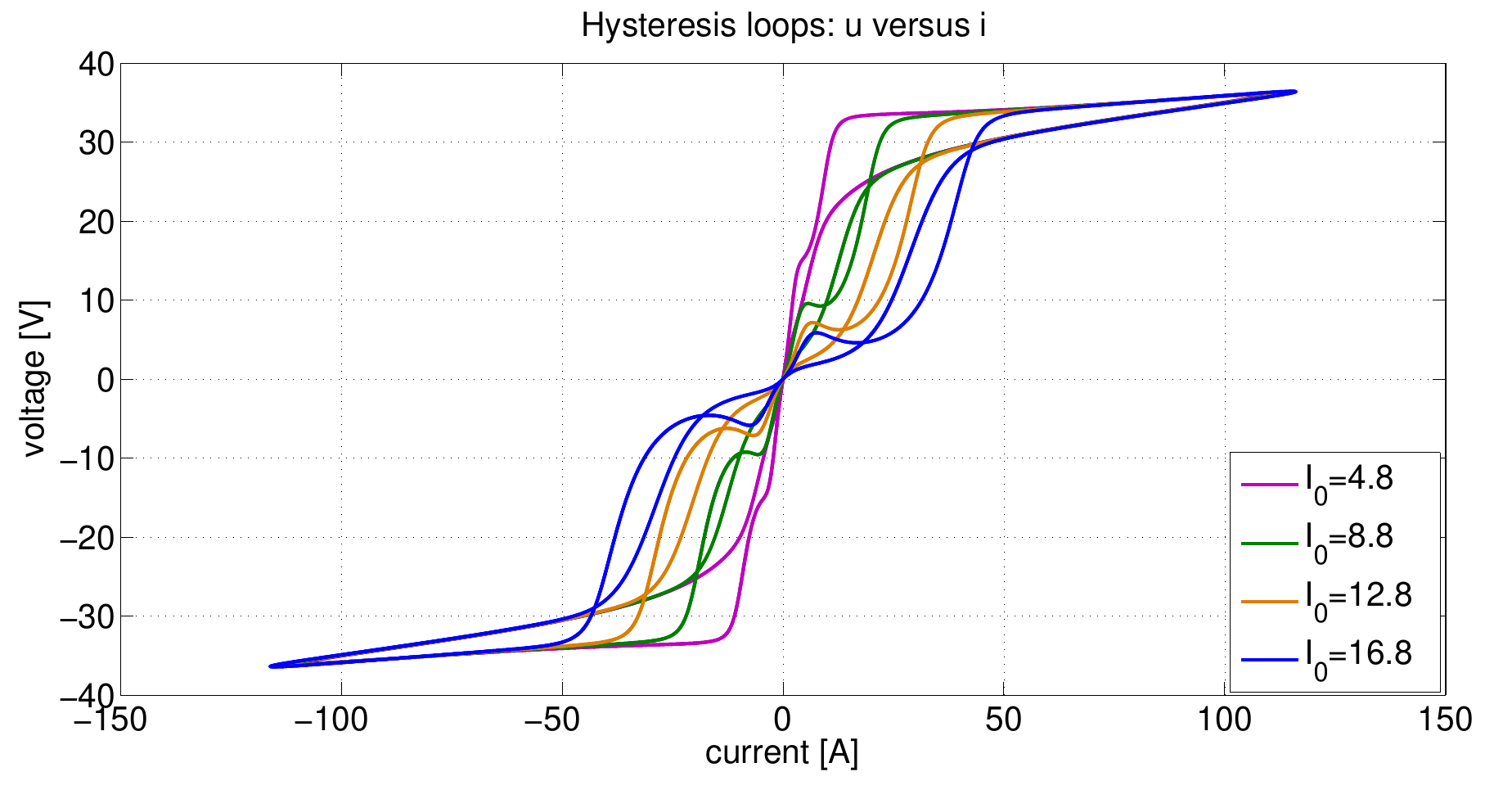}}
\subfigure[Motion around $(0,0)$ for $I_0=16.8$.]
{\label{Rys2b}\includegraphics*[height=1.85in,width=3.4in]{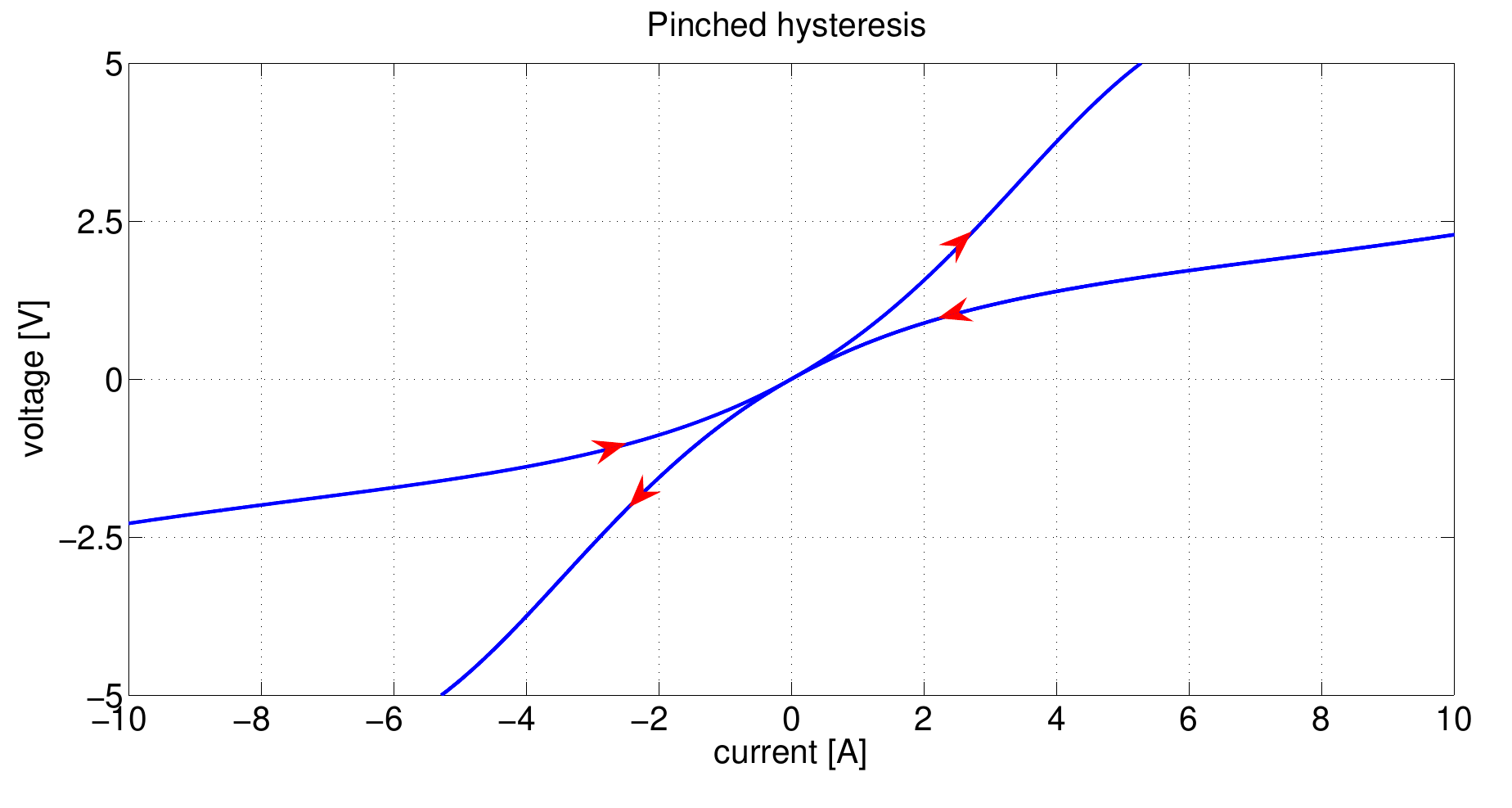}}
\caption{Various pinched hystereses.}
\label{Fig2} 
\end{center}
\end{figure}

\vspace{.15cm}
\noindent {\bf Remark 1:} The pinched property occurs at the origin, so $u\approx 0$ and $i\approx 0$. By using $u=i/g$ and the fact that the pinched property occurs at $i\rightarrow 0$, it is possible to show that $du/di$ in (\ref{1}),(\ref{2}) has {\bf one} positive value as $u\rightarrow 0$ and $i\rightarrow 0$, but $d^2u/di^2$ has {\bf two} different values (positive and negative) at the origin. The two opposite values of $d^2u/di^2$ occur half a period apart (see Fig.\ref{Rys2a}). This indicates that there are {\bf two} different trajectories of the $u$-$i$ characteristic at the origin, one that is concave up (with $d^2u/di^2>0$) and another that is concave down (with $d^2u/di^2<0$). See Fig.\ref{Rys2b} showing clearly {\bf two} tangential  trajectories with different concavity around the origin.

\vspace{.15cm}
\noindent {\bf Proof of fingerprint 1:} Notice that the pinched hysteresis occurs as  $i\rightarrow 0$. Thus, the Mayr model is in effect. Let $u(t_*)=0$ and $i(t_*)=0$ (see the second fingerprint below). We have $g(t_*)>0$. Also, since $u=i/g$, therefore we have $du/dt=[(di/dt)g-(dg/dt)i]/g^2=(di/dt)/g$ when $i=0$. This yields $du/di=1/g$. At $t=t_*$ we have $du/dt=(di/dt)/g(t_*)$, and since $g(t_*)>0$, therefore $du/di_{|t_*}>0$. Thus, the pinched hysteresis has slope $1/g(t_*)$ at $(u,i)=(0,0)$.

Now, we shall show that $d^2u/di^2$ is a two-valued quantity at $(u,i)=(0,0)$, that is $d^2u/di^2_{|t_*}$ and $d^2u/di^2_{|t_*+T/2}$ are of opposite signs. Using the fact that $i=0$ yields $du/di=1/g$, we obtain $d^2u/di^2=-(dg/di)/g^2=-(1/g^2)(dg/dt)/(di/dt)$. Note that the derivative $dg/dt$ is positive at $t=t_*$ and also from (\ref{2}) we have $di/dt_{|t_*}=(E_m/L)sin(2\pi ft_*)$, since $i(t_*)=0$ and $u(t_*)=0$. Thus, the Mayr model predicts that when the periodic, zero-average current $i(t)$ crosses the zero value at $t=t_*$, it is of a cosine type, with opposite signs of slope at $t=t_*$ and $t=t_*+T/2$. If $di/dt_{|t_*}>0$, then, half a period later we have $di/dt_{t_*+T/2}<0$. This yields $d^2u/di^2_{|t_*}<0$ and $d^2u/di^2_{|t_*+T/2}>0$. On the other hand, if  $di/dt_{|t_*}<0$, then, half a period later we have $di/dt_{t_*+T/2}>0$. This yields $d^2u/di^2_{|t_*}>0$ and $d^2u/di^2_{|t_*+T/2}<0$. This proves that the concavity of the trajectory (pinched hysteresis) is opposite at $t=t_*$ than at $t=t_*+T/2$. The trajectory moving in time along the $u$-$i$ characteristic is of different type of concavity when passing through $u=0$, $i=0$ every half of the period $T$, as illustrated in Figs.\ref{Rys2a} and \ref{Rys2b}. This completes the proof of the first memristive fingerprint of the hybrid Cassie-Mayr model. \hfill $\diamond$

 The facts that the slope $du/di$ has the same positive value at $t=t_*$ and at $t=t_*+T/2$ and opposite values of $d^2u/di^2$ at $t=t_*$ and $t=t_*+T/2$ yield the pinched hysteresis of type II, as discussed in \cite{Biolek1}-\cite{wm0}.

\vspace{.1cm}
\noindent {\bf Fingerprint 2:}  The $u$ and $i$ zero crossings occur at the same  instants.

\vspace{.15cm}
\noindent {\bf Remark 2:}  Since $u=i/g$, then, if $i(t_*)=0$ for some $t_*\ge 0$, then $u(t_*)=0$ when $G_{min}<g(t_*)< \infty$. Therefore, to avoid the symbol $0/0$, it suffices to show that the model (\ref{1}),(\ref{2}) yields the conductance $g(t_*)\ne 0$ . See Fig.\ref{Rys1d} for an illustration.

\vspace{.15cm}
\noindent {\bf Proof of fingerprint 2:} If $i(t_*)=0$ for some $t_*$, then, obviously, $i(t_*)<I_0$ and the hybrid arc model follows that of Mayr. Thus, from (\ref{3}) we have $g(t_*)=g_M(t_*)$ and the $g_M$ results from the Mayr model $g_M=G_{min}+i^2/P_0-\theta dg_M/dt$. When $i(t)\rightarrow 0$, then we have $g_M(t)\rightarrow G_{min}+Ce^{-t/\theta}$. Since the $g(t)$ must be greater or equal $G_{min}$ for small $i(t)$ (see a remark in section 1), therefore we must have that $C>0$. This yields $g_M(t_*)\ne 0$.\hfill $\diamond$

\vspace{.1cm}
\noindent {\bf Fingerprint 3:}   As $f\rightarrow \infty$, then the $u$-$i$ pinched hysteresis characteristic becomes that of a memoryless resistor, meaning that the $u$-$i$ graph is a single-valued one.

\vspace{.15cm}
\noindent {\bf Remark 3:} The third fingerprint can be proved, by analyzing the area enclosed by the hysteresis $u$-$i$ as $f\rightarrow \infty$. Notice that $f$ is the frequency of $E$ in (\ref{2}). We shall show that the enclosed area shrinks to zero as $f\rightarrow \infty$. This gives a single-valued relationship  between $u$ and $i$.

\vspace{.15cm}
\noindent {\bf Proof of fingerprint 3:} Since $E(t)=E_msin(2\pi ft)$ in (\ref{2}) and the fact that $u(t)$ and $i(t)$ are periodic, we can assume that 
\begin{equation}\label{5}
\begin{array}{rl}
u=&\sum_{k=1}^{\infty}[a_kcos(2\pi fkt)+b_ksin(2\pi fkt)]\\
i=&\sum_{k=1}^{\infty}[c_kcos(2\pi fkt)+d_ksin(2\pi fkt)]
\end{array}
\end{equation}
for some real numbers $a_k$, $b_k$, $c_k$ and $d_k$, $k=1, 2,\dots $.

The area, say $A$, enclosed by the pinched hysteresis loop $u$-$i$ over half period $t_*\le t<t_*+T/2$, $T=1/f$, equals 
\begin{equation}\label{eq30}
A=\oint u di=\int_{t_*}^{t_*+T/2}\!\!u(di/dt)dt.
\end{equation}
 Using (\ref{5}) and $di/dt=(E-u-Ri)/L$ from (\ref{2}) we obtain
\begin{equation}\label{6}
\begin{array}{rcl}
\!\!A&\!\!=\!\!&\frac{1}{L}\int_{t_*}^{t_*+T/2}\sum\limits_{k=1}^{\infty}\left [a_kcos(2\pi fkt)+b_ksin(2\pi fkt)\right ]\times \\ 
 & &\{\!E_msin(2\pi ft)\!-\!\!\sum\limits_{k=1}^{\infty}\!\left [a_kcos(2\pi fkt)\!+\!b_ksin(2\pi fkt)\right ]\\
 & &\hspace{1.5cm}-R\sum\limits_{k=1}^{\infty}\!\left [c_kcos(2\pi fkt)\!+\!d_ksin(2\pi fkt)\right ]\}dt.
\end{array}
\end{equation}
Notice that the right-hand side of (\ref{6}) contains integrals of various products of the cosine and sine terms. The integrals are computed over half of the period $T=1/f$, that is for $t_*\le t<t_*+T/2$. Such integrals can be computed according to the well-known formulas
\begin{equation}\label{7}
\int_{t_*}^{t_*+T/2}\!\!\!\!\!p_ksin(2\pi fkt)\times q_lsin(2\pi flt)dt=\Bigg \{\!\!\begin{array}{lll}\frac{p_kq_k}{4f}&\!\!for\!\!&k=l\\ 0 &\!\!for\!\!&k\ne l\end{array}
\end{equation}
\begin{equation}\label{8}\begin{array}{l}
\int_{t_*}^{t_*+T/2}\!\!p_kcos(2\pi fkt)\times q_lsin(2\pi flt)dt\\
\hspace{1cm}=\Bigg \{\begin{array}{lll}0&for&k=l\\ \frac{p_kq_ll[1-cos(\pi k)cos(\pi l)]}{2\pi f(l^2-k^2)}&for&k\ne l\end{array}\end{array}
\end{equation}
In addition, the same right-hand side holds true if we replace both sine terms by cosine terms in (\ref{7}).

Notice that by using the above integrals and the fact that $1/f$ is present in the non-zero right-hand sides in (\ref{7}) and (\ref{8}),  we obtain the right-hand side of  (\ref{6}) in the form of an infinite series with each term proportional to $1/f$, reciprocal of frequency. Thus, if $f\rightarrow \infty$, then the area of the pinched hysteresis decays to zero. This means that the $u$-$i$ characteristic becomes a single-valued one and the proof of the third fingerprint is complete. \hfill $\diamond$

\begin{figure}[b!]
\begin{center}
\subfigure[The $g$-$i$ loops for $f=\{0.4, 3, 5, 7, 9, 11\}$ kHz.]
{\label{Rys3a}\includegraphics*[height=1.85in,width=3.4in]{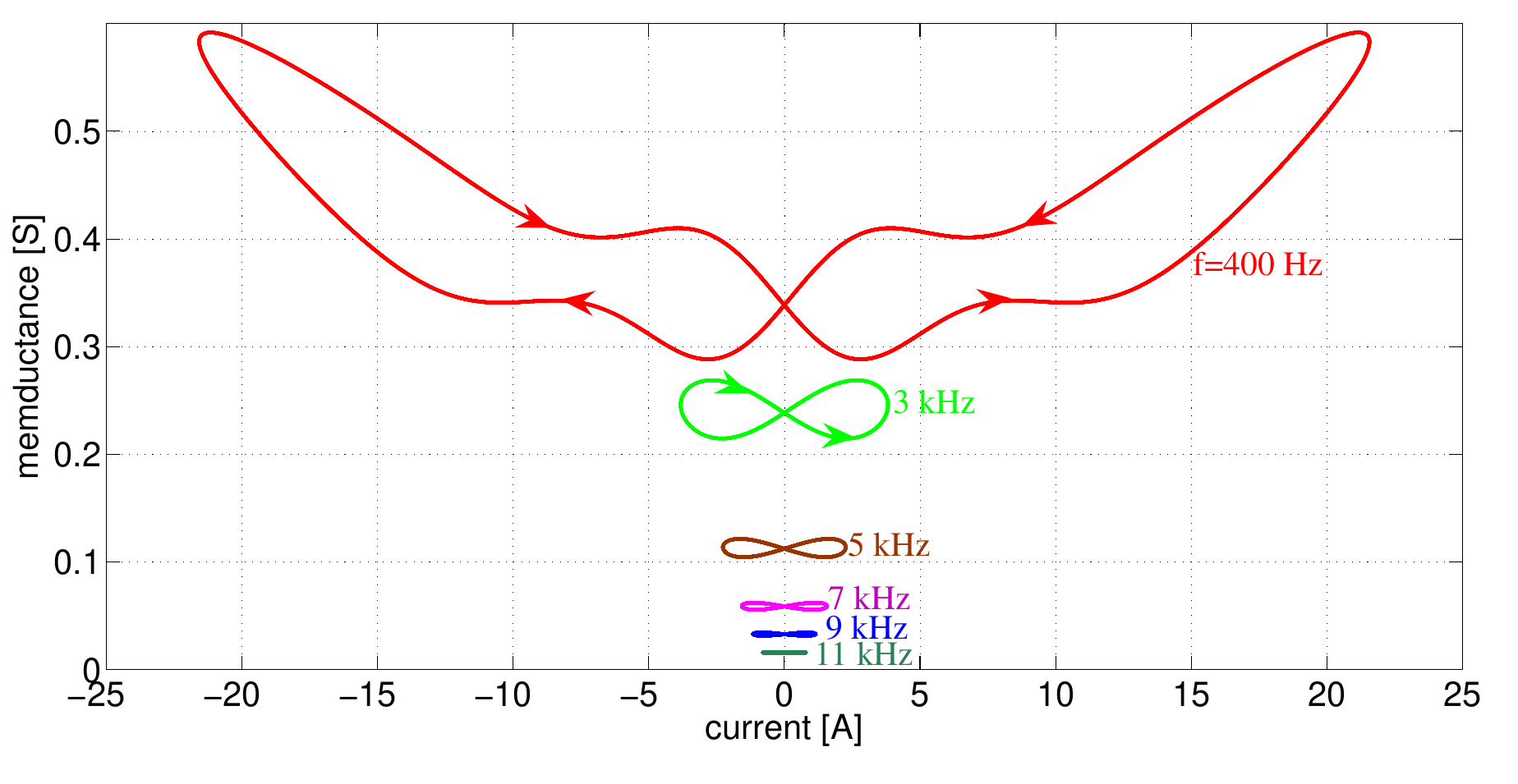}}
\subfigure[Enlargements of the $g$-$i$ loops for $f=\{5, 7, 9, 11\}$ kHz in Fig.\ref{Rys3a}.]
{\label{Rys3b}\includegraphics*[height=1.85in,width=3.4in]{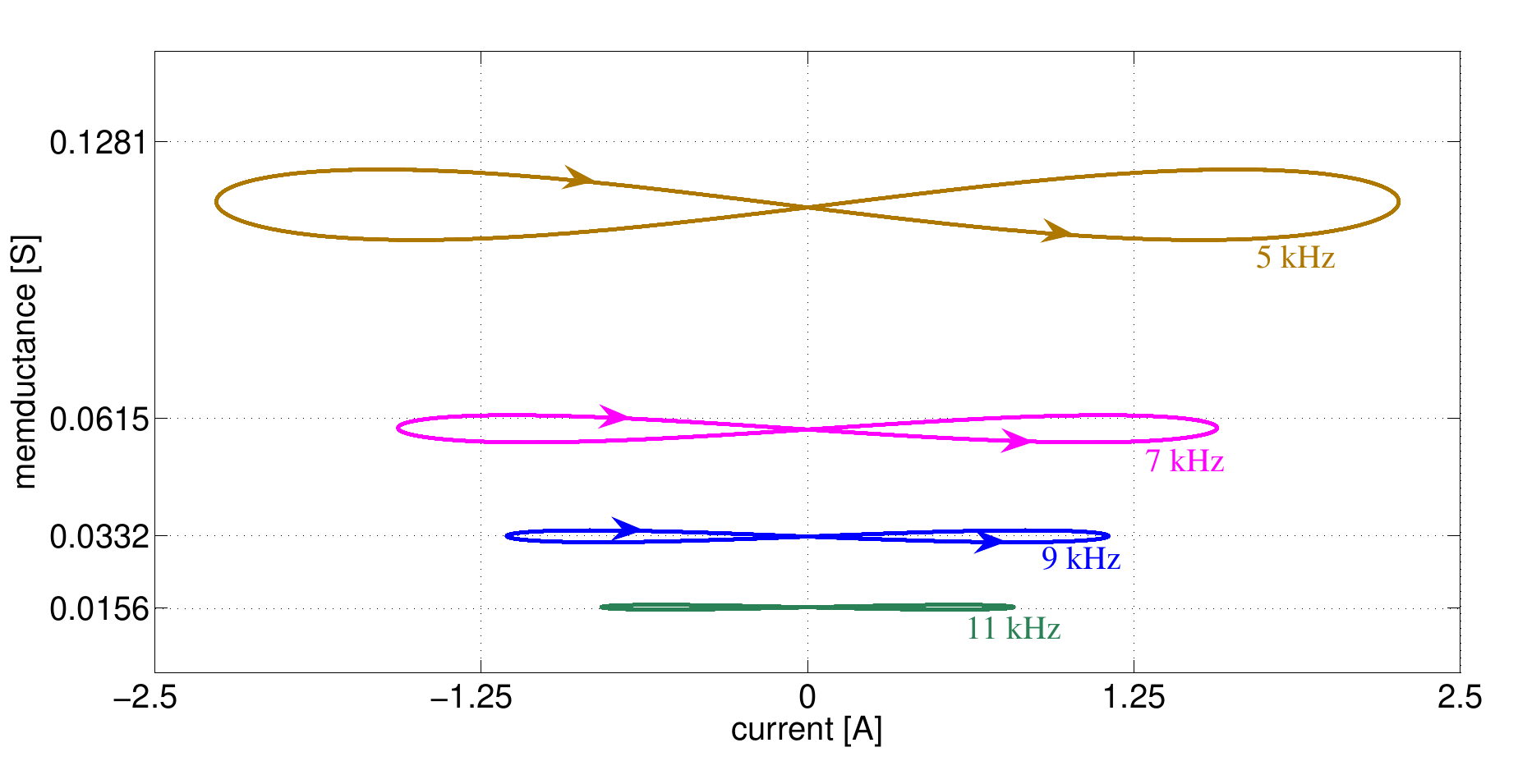}}
\caption{The $g$-$i$ loops for various values of $f$ in $E(t)$ and $\theta=2\cdot 10^{-4}$, $K=0.5$. Other parameters as given in Fig.1.}
\label{Fig3} 
\end{center}
\end{figure}

Fig.\ref{Rys1c} indicates that when the frequency $f$ increases, then the $v$-$i$ characteristic not only becomes closer to a single-valued one, but it becomes a linear with decreasing values of $g$ and $i$ (assuming fixed value of $E_m$). The $g$ is practically constant  in Fig.\ref{Rys1c} for a particular large value of $f$. Note that since $i<I_0$ for large $f$ values, therefore the Mayr model dominates in (\ref{1}). One can estimate the constant value of $g$ for large $f$, by using the well-known frequency formula for the Mayr model \cite{russian}. Namely, the sinusoidal current $i(t)=I_msin(2\pi ft)$ in the Mayr model yields $g(t)=G_{min}+\frac{I_m^2}{2P_M}\left \{ 1+\frac{cos(4\pi ft-\phi)}{\sqrt{1+16\pi ^2f^2\theta ^2}}\right \}$ with $\phi=tan^{-1}(4\pi f\theta)$. Thus $lim_{f\rightarrow \infty}g(t)=G_{min}+\frac{I_m^2}{2P_M}$. Fig.3 and and the associated Table 1 illustrate the use of the above limit in a simple numerical example.

\begin{table}
\caption{The $G_{min}+I_m^2/(2P_M)$ values for various frequencies.}
\label{tab0}   
\begin{center}    
\begin{tabular}{lll}
\hline \noalign{\smallskip}
 $f$ [kHz] & $I_m$ [A] &  $G_{min}+I_m^2/(2P_M)$ [S] \\
\noalign{\smallskip}\hline\noalign{\smallskip}
3 & 3.821&\hspace{1.3cm}0.3650\\
5 & 2.264 &\hspace{1.3cm}0.1281\\
7 & 1.568 &\hspace{1.3cm}0.0615\\
9 &  1.152&\hspace{1.3cm}0.0332\\
11&  0.790 &\hspace{1.3cm}0.0156\\
\noalign{\smallskip}
\hline
\end{tabular}
\end{center}
\end{table}

Comparing Fig.\ref{Rys3b} and Table \ref{tab0} it is easy to notice that, with the increased frequency $f$, the $G_{min}+I_m^2/(2P_M)$ are indeed very good estimates of the \emph{almost} constant $g$ values. The last four values in the third column in Table \ref{tab0} are marked on the vertical axis in Fig.\ref{Rys3b}.

\section{Variation of hystereses with parameters}
\begin{figure}[t!]
\begin{center}
\subfigure[Hysteresis loops for various values of $K$.]
{\label{Rys4a}\includegraphics*[height=1.85in,width=3.4in]{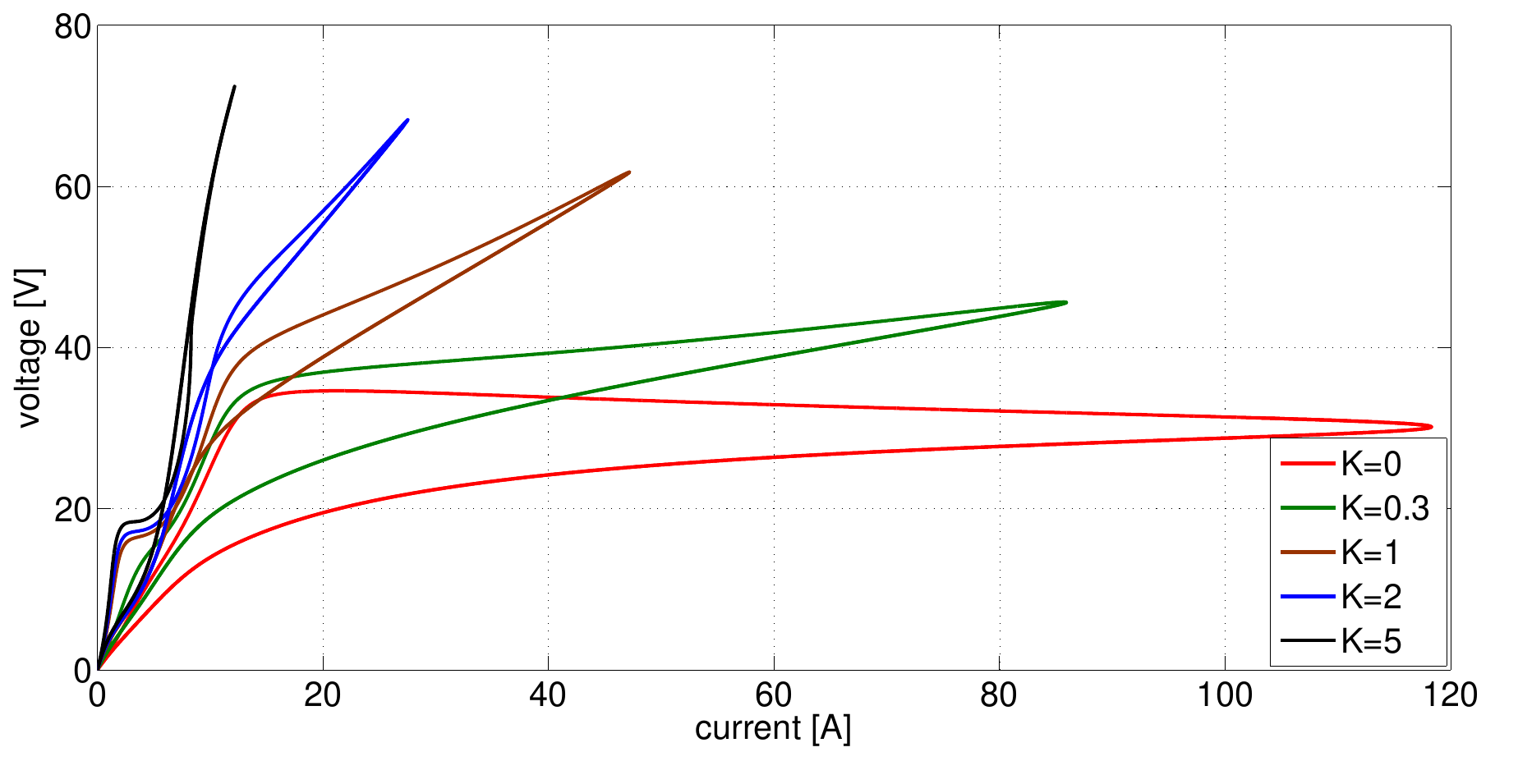}}
\subfigure[Hysteresis loops for various values of $L$.]
{\label{Rys4b}\includegraphics*[height=1.85in,width=3.4in]{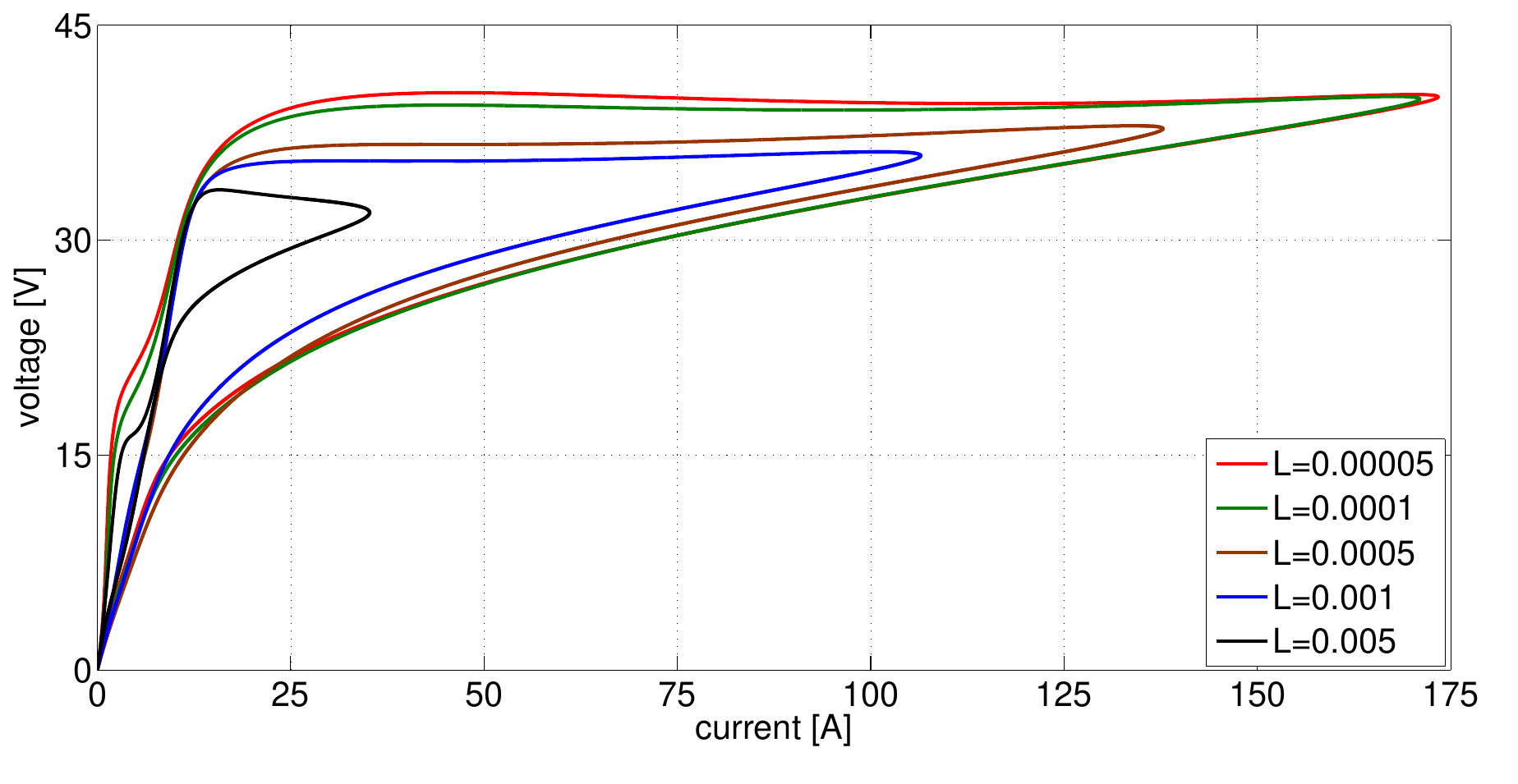}}
\subfigure[Hysteresis loops for various values of $U_C$.]
{\label{Rys4c}\includegraphics*[height=1.85in,width=3.4in]{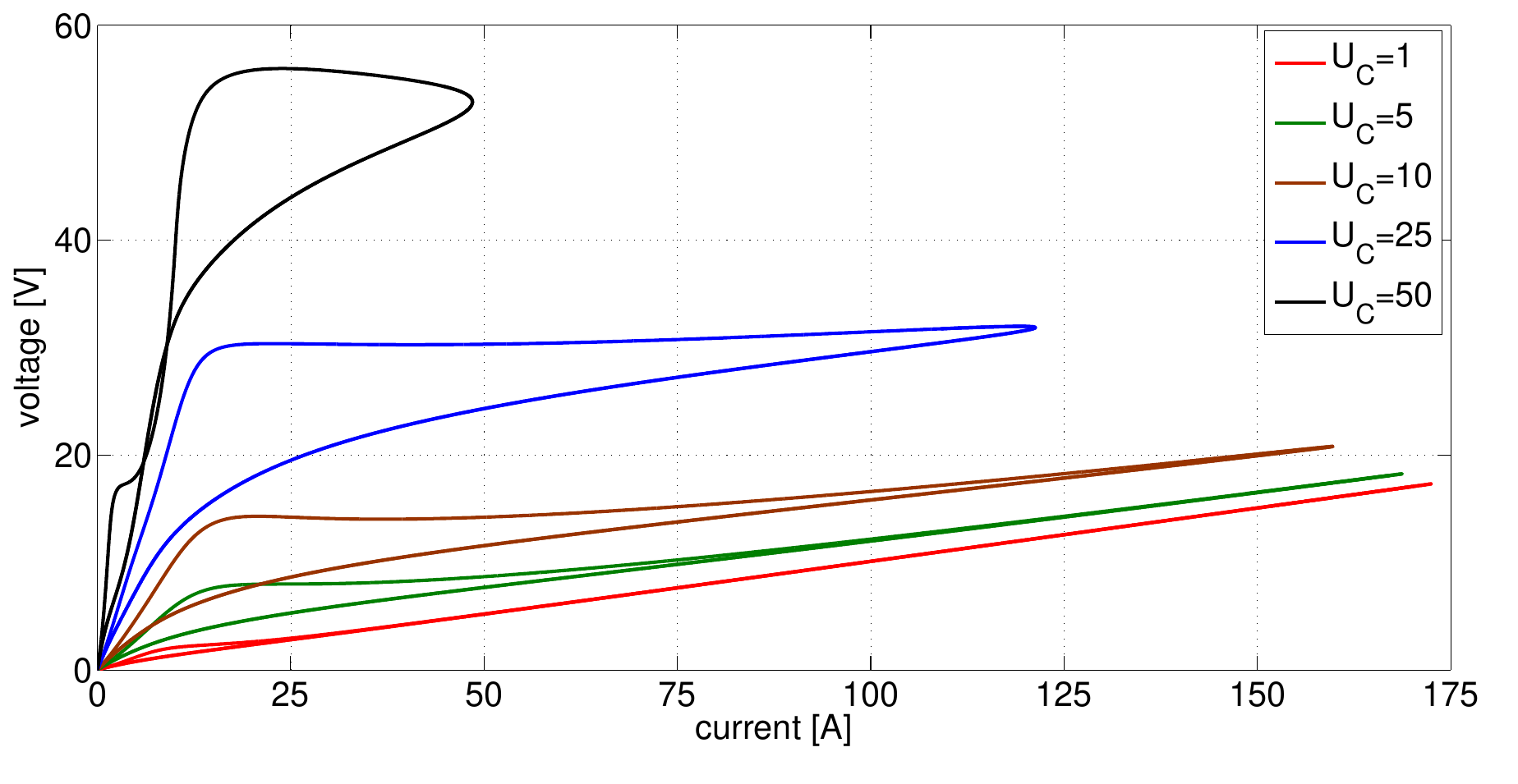}} 
\caption{Impact of parameters $K$, $L$ and $U_C$ on the pinched hysteresis loops of the Cassie-Mayr model (\ref{1}),(\ref{2}). The \emph{ode45} solver from Matlab with \emph{abserr}=\emph{relerr}=$10^{-10}$ was used.}
\end{center}
\label{Fig4} 
\end{figure}

Figs. \ref{Rys1b} and \ref{Rys2a} show various shapes of the hysteresis loops of (\ref{1}),(\ref{2}) when the frequency $f$ and current $I_0$ change, respectively. Other parameters in the Cassie-Mayr model impact the hystereses, too. Fig. \ref{Rys4a}, \ref{Rys4b} and \ref{Rys4c} illustrate such an impact when the parameters $K$, $L$ and $U_C$ vary, respectively. The constant parameters in all three figures were:  $\theta=4~\times~10^{-4}$, $G_{min}=10^{-8}$, $I_0=4.8$, $P_M=20$, $R=0.2$, $f=50$ and $E_m=75$. In addition, $U_C=30$, $L=10^{-3}$ and $K=\{0, 0.3, 1, 2, 5\}$ in Fig.\ref{Rys4a}. Also, $U_C=30$, $K=0.1$ and $L=\{5\cdot 10^{-5},10^{-4},5\cdot 10^{4},10^{-3},5\cdot 10^{-3}\}$ in Fig.\ref{Rys4b}. Finally, $K=0.1$, $L=10^{-3}$ and $U_C=\{1,5,10,25,50\}$ in Fig.\ref{Rys4c}. The hysteresis loops are shown in the first quadrant only (positive voltage and current values). Symmetric graphs exist in the third quadrant with negative voltage and current values.

\section{Conclussion}
The seemingly distant and disjoint areas of electric arcs from welding, electric furnaces and circuit breakers on one side and  memristors from nanometer electronics on the other side have been linked together through their identical mathematical properties (fingerprints). It was shown through mathematical analysis that the hybrid Cassie-Mayr model of electric arcs has all the fingerprints of memristors, passive nonlinear nanoelements with memory.  By linking the electric arcs with memristors one can now apply various techniques and methods from the nanoscale electronics (i.e. to  analyze  energy and power \cite{Paper10},\cite{wm1},\cite{wm2}) to the nonlinear plasma phenomena in electric arc furnaces, circuit breakers and welding processes \cite{pap4}-\cite{pap7}.

\section{Acknowledgement}
The author would like to thank Prof. Z. Trzaska from Warsaw (Poland) for his discussion on the topic of electric arcs.


\end{document}